\documentclass[twocolumn]{article}
\usepackage[latin1]{inputenc}  
\usepackage{graphicx}

\title{\footnotesize{To appear in Journal of Physical Chemistry A (2009) - "Chemistry: Titan Atmosphere"} \\  \LARGE \textbf{DNA nucleobase synthesis at Titan atmosphere analog by soft X-rays}}

\author{Sergio Pilling\footnote{sergiopilling@yahoo.com.br} (PUC-Rio); Diana P. P. Andrade (PUC-Rio) \\ Álvaro C. Neto (Unicamp), Roberto Rittner (Unicamp), Arnaldo Naves de Brito (LNLS)}

\begin{document}
\maketitle

\begin{abstract}

Titan, the largest satellite of Saturn, has an atmosphere chiefly made up of N$_2$ and CH$_4$ and includes traces of many simple organic compounds. This atmosphere also partly consists of hazes and aerosols particles in which during the last 4.5 gigayears have been processed by electric discharges, ions, and ionizing photons, being slowly deposited over the Titan surface.
In this work we investigate the possible effects produced by soft X-rays (and secondary electrons) in Titan aerosol analogs in an attempt to simulate some prebiotic photochemistry. The experiments have been performed inside a high vacuum chamber coupled to the soft X-ray spectroscopy (SXS) beamline at the Brazilian Synchrotron light Source (LNLS), Campinas, Brazil. \textit{In-situ} sample analysis were performed by a Fourier transform infrared spectrometer (FTIR). The infrared spectra have presented several organic molecules including nitriles and aromatic CN compounds. After the irradiation, the brownish-orange organic residue (tholin) were analyzed \emph{ex-situ} by gas chromatographic (GC-MS) and nuclear magnetic resonance ($^1$H NMR) techniques, revealing the presence of adenine (C$_5$H$_5$N$_5$), one of the constituents of DNA molecule. This confirms previous results which showed that the organic chemistry onto Titan surface can be very complex and extremely rich in prebiotic compounds. Molecules like these in the early Earth have found a place to allow life (as we know) to flourish.
\end{abstract}

\section{Introduction}

Titan is the largest moon of Saturn, and also one of the largest moons in the Solar System. Discovered by Christian Huygens in 1655, it is larger in diameter than the planets Mercury and Pluto. This alone would make Titan an intriguing target for exploration. Titan was visited by the Voyager 1 spacecraft in 1980, and more recently (October 2004) by the Cassini/Huygens mission to Saturn. The Titan atmosphere's is chiefly made up of N$_2$ and CH$_4$ with traces of many small organic molecules (e.g. hydrocarbons and nitriles). With its dunes, lakes, channels, mountains, and cryo-volcanic features\cite{LeCorre2008}, Titan is an active place that resembles Earth, with methane playing the role of water and ice that of silicates\cite{Grasset96, mitri2008}. Titan atmosphere also partly consists of hazes and aerosols particles which shroud the surface of this satellite, giving it a reddish appearance. As a consequence of its high surface atmospheric pressure ($\sim$1.5 bar) the incoming solar ultraviolet (UV) and soft X-ray photons are mostly absorbed. As a consequence, only low amounts of energetic photons reach the surface. However, during the last 4.5 gigayears, the photolyzed atmospheric molecules and aerosol particles have been deposited over the Titan surface composed by water-rich ice (80-90 K) delivered by comets. As pointed by Griffiths and coworkers\cite{Griffith03} this process produced in some regions a ten meter size, or even higher, layers of organic polymer also known as "tholin".

The term tholin was coined about thirty years\cite{sagan79} ago to describe the products obtained by the energetic processing of mixtures of gases abundant in the cosmos, such as CH$_4$, N$_2$, and H$_2$O. Tholin comes from the Greek, meaning "muddy", an apt description for the brownish, sticky residues (general formula CxHyNz) formed by such experiments. These experiments, using either electrical discharges or ultraviolet irradiation, are the natural extensions of the well-known Miller-Urey experiment\cite{miller59}. While the
Miller-Urey experiment focused on an atmosphere meant to be like that of the
early Earth, Sagan and others attempted to simulate the atmospheres of other
planets and moons in the Solar System, such as Titan and Triton\cite{McDonald94, sagan79, Khare73, moore2003}. When placed in liquid water, some of tholins's compounds (water soluble) have been shown to produce oxygenated organic species\cite{Neish08}.

The investigation of Titan tholins produced by electric discharges, UV photolysis and radiolysis have been extensively performed\cite{Khare84, Khare86, McDonald94, Coll98, moore2003, Imanaka04, Tran05, Tran03, Imanaka07, McGuigan06}. However the photochemistry promoted by soft X-rays on primitive/extraterrestrial atmospheres as well on Titan atmosphere analog were poorly analyzed. In this work we investigate the effects produced by the interaction of soft X-rays and secondary electrons on Titan aerosols analogs containing a solid mixture at about 15 K made mainly by N$_2$ (95\%), CH$_4$ (5\%) and traces of water and CO$_2$. In Section 2, we present briefly the experimental setup and the analysis methods utilized. The results and discussion are given in Section 3. Final remarks and conclusions are given in Section 4.

\section{Experimental Setup and Methods}

\subsection{Irradiation of Titan aerosol analog and \textit{in-situ} analysis}

In an attempt to simulate the photochemistry process ruled out by soft X-rays on Titan atmosphere analog, we use the facilities of the the Brazilian Synchrotron Light Laboratory (LNLS) located in Campinas/Brazil. The experiments were performed inside a high vacuum chamber coupled to the soft X-ray spectroscopy (SXS) beamline employing a continuum wavelength beam from visible to soft X-rays with a maximum flux between 0.5-3 keV range.

A gas mixture simulating the Titan atmosphere (95\% N$_2$, 5\% CH$_4$) was
continuously deposited onto a polished NaCl substrate previously cooled at 14 K in a high vacuum chamber and exposed to synchrotron radiation up to 73 h. The atmosphere inside the chamber was monitored by quadrupole mass analyzer (Residual Gas Analyzer - RGA; PrismaTech 100). During the irradiation the sample temperature had a small increase and stabilize at about 15 K.

\begin{figure}[!t]
 \centering
\includegraphics[scale=0.40]{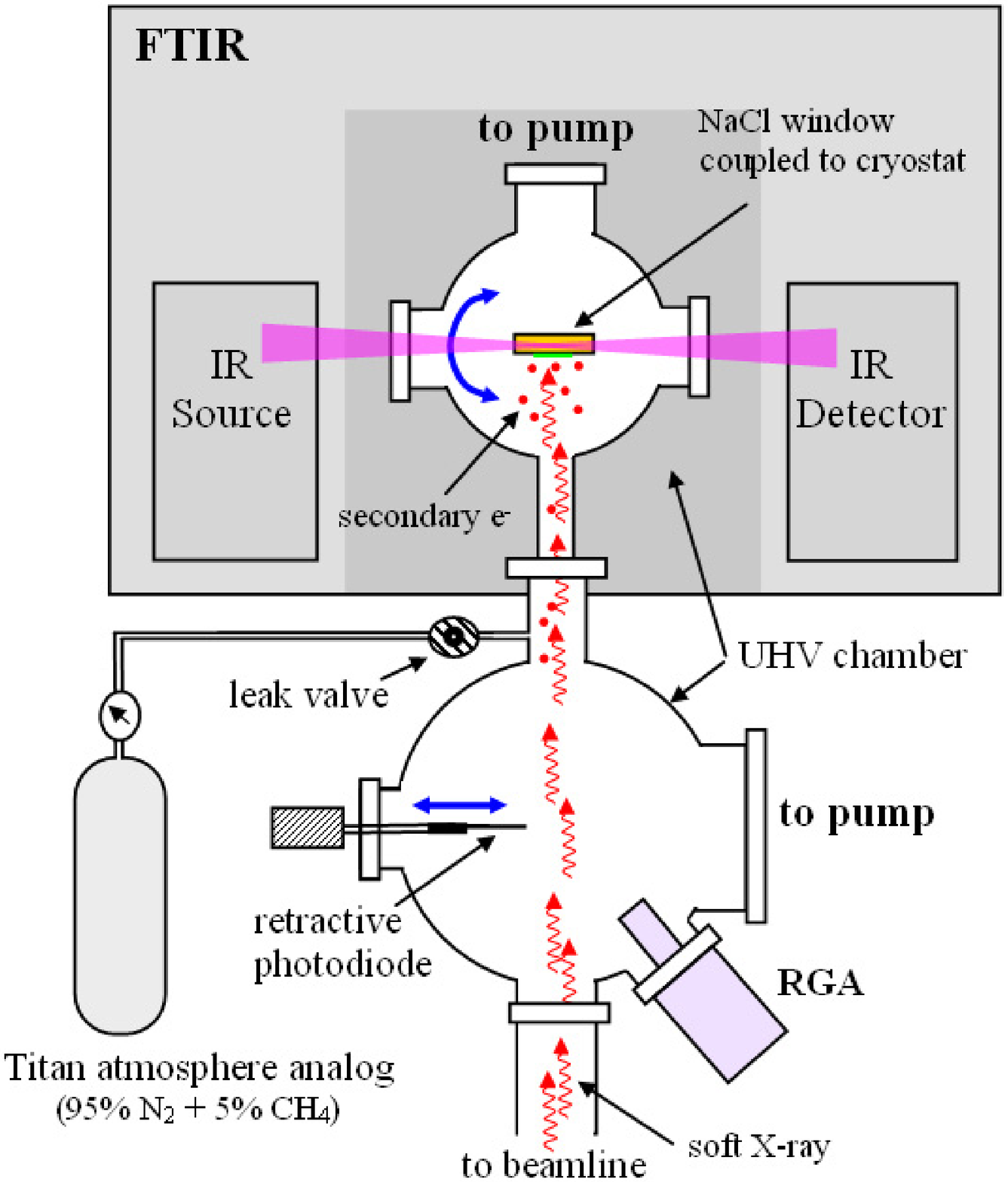}
\caption{A schematic diagram of the experimental set-up. The photon beam hits perpendicularly the a NaCl crystal at the same time that the dosing is occurring. During the dosing the substrate is turned to the retractive inlet sample. For irradiation the substrate with the ice sample is turned 180$^{\circ}$ to the photon  beam. After each irradiation dose the target is rotated by 90$^{\circ}$ for FTIR analysis. } \label{fig:diagram}
\end{figure}

\emph{In-situ} sample analysis were performed by a Fourier transform infrared spectrometer (FTIR-400, JASCO Inc.) coupled to the experimental chamber. The infrared beam from FTIR and the synchrotron beam intercept perpendicularly at the sample. The infrared (IR)
transmission spectra were obtained rotating the substrate/sample by 90 degrees after each radiation dose. Infrared spectra of non-irradiated samples were taken at the beginning and at the end the experiments and were compared. A schematic representation of the experimental chamber is shown in Fig.~\ref{fig:diagram}.

During the deposition/irradiation phase the pressure on the chamber was about
$2\times10^{-6}$ mbar. This allows a continuous flux of non-irradiated molecules (about 1-2 monolayers per second) which can react with the photoproducts trapped on the icy surface. The chamber base pressure was about $8\times10^{-8}$ mbar. At this pressure a layer of water molecules and CO$_2$, due to residual gas, was deposited on the substrate roughly after every minute. This fraction of residual water molecules (about 1-5\% of the icy sample) and carbon dioxide supplies an oxygen source for the photochemical reactions, simulating thus a possible cometary delivery (or other water sources) in Titan.

The beamline details can be found elsewhere\cite{Abate99}. The continuum wavelength photon distribution (white beam mode) was obtained by placing the beamline monocromator out from the line of sight allowing photons from near IR up to 4 keV reach the experimental chamber. The determination of the photon flux at sample was done by the following procedure: 1) measurement the UV photon flux using a narrow filter (3.2-3.4 eV) by a photosensitive diode (AXUV-100, IRD Inc.) coupled to the experimental chamber; 2) scaling the theoretical beamline transmission flux, obtained employing the \texttt{XOP/SHADOWVUI} ray-tracing code software\footnote{\texttt{http://www.esrf.eu/computing/scientific/xop2.0/}} to the measured UV photon flux.

\begin{figure}[!t]
 \centering
\includegraphics[scale=0.25]{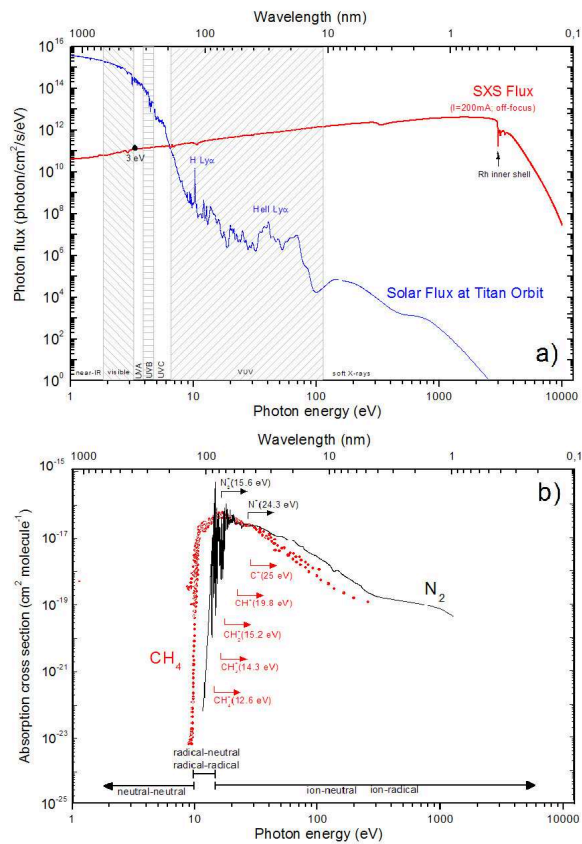}
\caption{a) Comparison between SXS beamline photon flux and the solar flux
at Titan orbit\cite{Gueymard04}. The measured narrow band photon flux at 3 eV is also indicated. b) Absorption coefficient of Titan atmosphere major
constituents. The appearances potential for the main ionic photodissociative
channels are also indicated. See details in text.} \label{fig-SXSbeam}
\end{figure}

The computed SXS beamline photon flux as a function of photon energy can be seen at Fig.~\ref{fig-SXSbeam}a in comparison with the solar photon flux at Titan orbit\cite{Gueymard04}. The integrated photon flux at the sample, which are ruled out mainly by the  photons between 0.1 to 5 keV, was about $\sim$ 10$^{16}$ photons cm$^{-2}$ s$^{-1}$ or roughly 10$^{7}$ erg cm$^{-2}$ s$^{-1}$. The solar integrated photon flux between 0.1 to 5 keV at Titan orbits correspond to $\sim$ 10$^{7}$ photons cm$^{-2}$ s$^{-1}$, a value about 9 orders of magnitude lower than achieved by SXS beamline. Therefore each hour of sample exposure to soft-X rays at SXS beamline corresponds to roughly 10$^5$ years to solar X-rays photons.

The beamline entrance and exit slit were completely opened during the experiments to allow the maximum intensity of the beam line. In an attempt to increase the beam spot at the sample, the experimental chamber was placed at about 1.5 m away from the beam line focus. With this procedure the measured beam spot at the sample was about 0.6-0.5 cm$^2$. Fig.~\ref{fig-SXSbeam}b presents the absorption coefficient of the major constituents of the Titan atmosphere. The appearance potentials for the main ionic photodissociative channels of N$_2$ and CH$_4$ are indicated. The photochemistry regime bellow 10 eV is governed by neutral-neutral. For energies between approximately 10 to 14 eV the chemical pathway involves neutral-radical and radical-radical. For energies higher than 15 eV the reaction involving ionic species rule out the photochemistry. In the case of soft X-rays ($\sim$ 0.1-10 keV) the produced secondary electrons also become an important route for molecule processing.

We observe a small enhancement of H$_2$ on the residual gas during the irradiation attributed to the photodissociation of CH$_4$ on gas phase. This indicates that some gas phase photochemistry (and also due to secondary electrons) were also occurring inside the chamber.

\subsection{\textit{Ex-situ} chemical analysis of the organic residue}

After the irradiation phase, the sample was slowly heated up to room temperature and another set of IR spectra were collected to follow the chemical changes promoted by thermal heating. A similar heating could be achieved locally at Titan surface during a comet impact or volcanism events. Next, the chamber was filled with dry nitrogen up to atmosphere pressure. The NaCl substrate, with the brownish-orange organic residue (tholin), were disconnected from the sample holder, conditioned into a sterile vial and sent to chromatographic and NMR analysis.

The general protocol for the chemical analysis of the amino acids with gas chromatography coupled to mass spectrometry (CG-MS) is described elsewhere\cite{Abe96,Nuevo2007}. In this method, the amino acids are derivatizated to volatile compounds which allows their separation in the gas chromatography column. The residues were first extracted from their NaCl window with $3\times30$ $\mu$L of H$_2$O using a sterilized vial. The water was evaporated by placing the vial in a desiccator at reduced pressure ($\sim$ 10 mbar). Once the water had totally evaporated, the sample was hydrolysed in 300 $\mu$L of 6M HCl and kept for 24 h in an oil bath maintained at 110 $^{\circ}$C. During this step, peptides and/or amino acids precursors (if they are present) are converted into free amino acids\cite{Nuevo2007}. Then, the HCl solution was evaporated in the desiccator at reduced pressure. The sample was then dissolved in 50 $\mu$L of 0.1M HCl and 25 $\mu$L of an ethanol:pyridine = 3:1 mixture and 5 $\mu$L of ethyl chloroformate (EtOCOCl) were added to the sample to derivatize the carboxylic acid and amino groups. The vial was shaken vigorously and 15 $\mu$L of chloroform was added. Next, the vial was shaken again to extract the derivatives into the organic phase.

The extract was finally injected (1 $\mu$L) directly into the gas chromatography Rtx$^{R}$-1 CG-MS system equipped with a 30 m stationary dimethyl polysiloxane phase column (0.25 m inner diameter; Restek). Splitless injections were performed, with an oven temperature programmed to 0 min at 50 $^{\circ}$C, and heated with 10 $^{\circ}$C min$^{-1}$ to 90 $^{\circ}$C, 2 $^{\circ}$C min$^{-1}$ to 110 $^{\circ}$C, and 10 $^{\circ}$C min$^{-1}$ to 180 $^{\circ}$C, where it was kept constant for 21 min. Helium was used as a carrier gas with a constant flow of 1.5 mL min$^{-1}$. The irradiated sample chromatograms were recorded in the more sensitive single-ion monitoring (SIM) mode of the mass spectrometer.

The NMR spectra were obtained on a Bruker DPX 250 spectrometer equipped with an inverse 5 mm probe, operating at 250.13 MHz for $^1$H NMR. Spectra of both samples, from derivatization (described above), were taken at 300 K and referenced to Me$_4$Si. For the sample from the experiment in LNLS 25k transients were used, while for the authentic sample 2k transients were performed.

\section{Results}

\subsection{FTIR}

\begin{figure}[!th]
 \centering
\includegraphics[scale=0.23]{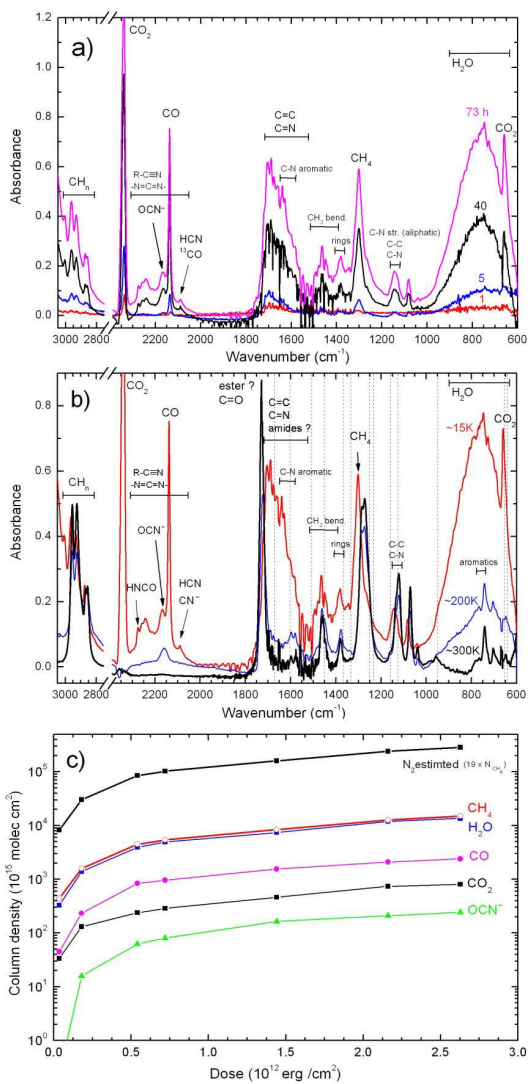}
\caption{a) FTIR spectra of the organic residue produced by the
irradiation of condensed Titan atmosphere analog under $\sim$ 15 K NaCl surface at different exposure times (1 h $\sim 3 \times 10^{10}$ erg cm$^2$). b) Comparison between FTIR spectra of 73 h irradiated sample at 15, 200 and 300K. The vertical dashed lines indicate the frequency of some vibration modes of crystalline adenine\cite{LappiFranzen04}; c) Molecular column density of the most important species observed on the ice as a function of irradiation dose. } \label{fig-FTIR}
\end{figure}

\emph{In-situ} FTIR spectra, from 3000 to 600 cm$^-1$, of the Titan tholin produced by the soft X-rays irradiation of condensed Titan atmosphere analog are given in Fig.~\ref{fig-FTIR}a and \ref{fig-FTIR}b. The upper panel presents IR spectra of the sample at $\sim$ 15 K at different irradiation doses up to 73 h. The molecular species related to the main IR feature are indicated. The narrow peak at about 2100 cm$^{-1}$ is the CO stretching mode ($\nu_1$). The broad feature 800 cm$^{-1}$ is the vibration mode ($\nu_L$) of water molecules. The CH$_4$ deformation mode ($\nu_4$) is observed around 1301 cm$^{-1}$. During the irradiation several features associated with nitriles (2100-2300 cm$^{-1}$), CH$_n$ (2800-3000 cm$^{-1}$), and aromatic molecules are also observed.

Fig.~\ref{fig-FTIR}b presents a comparison between the IR spectra at the maximum dose obtained after 73hs of exposure to soft X-rays ($\sim$ 2.7 $\times 10^{12}$ erg cm$^{-2}$) at three different temperatures: 15, 200 and 300 K. The well defined infrared bands associated with vibrational modes of newly produced nitriles (2100-2300 cm$^{-1}$) are still observed at around 200 K. However, at higher temperature this feature is not observed, a consequence of the complete evaporation of these kinds of nitriles. The organic residue at room temperature, present strong bands at 2800-3000 cm$^1$ associated with non-volatile hydrocarbons (CH$_n$), a intense and sharp peak at $\sim$1720 cm$^{-1}$, possibly attributed to C=O mode of esters, and several other unidentified features at 1450, 1375, 1290, 1140, 1070 cm$^{-1}$. Some of these bands could be due to C-N aromatic and rings\cite{Imanaka04}. The vertical dashed lines indicate the location of some vibration modes of pure crystalline adenine (C$_5$H$_5$N$_5$)\cite{LappiFranzen04}. A direct comparison between these frequencies and the Titan tholin infrared spectrum at 300 K has not shown a strong evidence of adenine molecules and only a tentative identification was possible (small peaks).

The variation of the column density of the abundant molecules observed during the irradiation of Titan atmosphere analog by soft X-rays as a function of dose is shown at Fig.~\ref{fig-FTIR}c. The lines were employed only to guide the eyes. The molecular column density was determined from the relation between the integrated absorbance, $Abs_{\nu}$ (cm$^{-1}$), of a given vibration mode with frequency $\nu$ in IR spectra and its respective band strength, $A$ (cm molec$^{-1}$):
\begin{equation}
N=\frac{1}{A} \int \tau_{\nu} d\nu = \frac{2.3}{A} \int Abs_{\nu} d\nu \quad \textrm{[cm$^{-2}$]}
\end{equation}
where $\tau_{\nu} = \ln (I_0/I) = 2.3 Abs_{\nu}$ is the optical depth (since $Abs_{\nu} = \log(I_0/I$)) and I$_0$ and I are the original and the attenuated infrared beam detected by the spectrometer, respectively. The vibrational features and its infrared absorption coefficients (band strengths) for the analyzed molecules in this work are given in Table~\ref{tab:A}.

\begin{table}[!htb]
\begin{center}
\caption{Infrared absorption coefficients (band strengths) used in the column density calculations for the observed molecules.} \label{tab:A}
\setlength{\tabcolsep}{2pt}
\begin{tabular}{ c c c c }
\hline \hline
Frequency ($\nu$)  & Assignment  & Band strength (A)      & Ref. \\
(cm$^{-1}$)       &             &   (cm molec$^{-1}$)  &           \\
\hline
2342           & CO$_2$ ($\nu_3$)   & 7.6$\times 10^{-17}$   & ~\cite{Gerakines1995} \\
$\sim$ 2234    & N$_2$O ($\nu_1$)     & 5.2$\times 10^{-17}$   & ~\cite{Wang01}  \\
$\sim$ 2165    & OCN$^-$ ($\nu_3$)    & 4$\times 10^{-17}$     &  ~\cite{Hendecourt86} \\
2139           & CO ($\nu_1$)         & 1.1$\times 10^{-17}$   &  ~\cite{Gibb 2004}  \\
1115-1065      & NH$_3$ ($\nu_2$)     & 1.2$\times 10^{-17}$   & ~\cite{Kerkhof99}  \\
$\sim$ 800     & H$_2$O ($\nu_L$)     & 2.8$\times 10^{-17}$   &  ~\cite{Gibb 2004}  \\
\hline \hline
\end{tabular}
\end{center}
\end{table}

The column density of frozen N$_2$ was estimated to be about 19 times the column density of CH$_4$ (from a mixture of 95\% N$_2$ and 5\% CH$_4$). For this assumption we also suppose that both molecules have approximately the same sticking coefficient and dissociation cross section. The amount of water and methane are virtually the same in the experiment. The abundance of CO$_2$ is about 10-20 times lower than water (roughly the same ratio observed in comets). The fraction of CO in the residual gas is very low, so the CO observed in the IR spectra is mainly due to the processing of CO$_2$. Initially the CO abundance is virtually zero (not shown in the Fig.~\ref{fig-FTIR}c) but just after the first hour of exposure to the soft-X rays, a fraction of the frozen CO$_2$ is converted in CO.

After 15 h of irradiation, about a dose of $5\times 10^{11}$ erg cm$^2$, the ratio CO/CO$_2$ reaches a constant value around 3.5, mainly due to the reverse processing of CO to from CO$_2$. In this situation the number of CO produced by CO$_2$ is equal the amount of CO$_2$ produced by CO. The CO column density still increase due to the continuous deposition of CO$_2$ on the cryogenic NaCl substrate. The column density of one reactive CN compound, the cyanate ion (OCN$^-$) is another example of the new formed species due to the processing of frozen Titan atmosphere analog. Moore and Hudson\cite{moore2003} in a similar experiment involving proton irradiation and UV photolysis, have also observed the formation of OCN$^-$. As in the present work this species was still detected at 200 K, evaporating at higher temperatures.

\subsection{GC-MS and NMR}

\begin{figure}[!th]
 \centering
\includegraphics[scale=0.16]{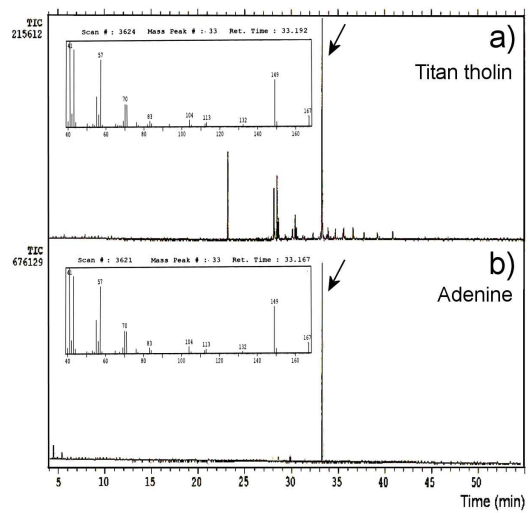}
\caption{Total-ion current (TIC) chromatogram obtained after the derivatization
process of the Titan tholin produced by soft X-rays (a) and adenine standard (b). The inset figures are the mass fragmentation of the sample and the adenine derivative at retention time of around 33.18 min (arrows).} \label{fig-GCMS}
\end{figure}

Following the methodology described above after the sample heating to room temperature the organic residue was derivatizated to volatile compounds, and one portion was injected directly into the gas chromatograph. The total ion-current (TIC) chromatogram obtained for sample is given in Fig.~\ref{fig-GCMS}a and show an intense signal around 33.2 min. The comparison between the Titan tholin sample and the literature data or the standard amino acids derivatives prepared in our laboratory, has not showed any amino acids formation. However, the preparation of adenine derivative using the same procedure described above and the analysis in the CG-MS showed a signal at 33.16 min in the chromatogram. The mass fragmentation of the adenine derivative is identical to the sample obtained in the Titan experiment (Fig.~\ref{fig-GCMS}b). These values of retention time and mass fragmentation confirm the production of adenine from the irradiation of Titan atmosphere analog by synchrotron soft-Xrays.

\begin{figure}[!th]
 \centering
\includegraphics[scale=0.20]{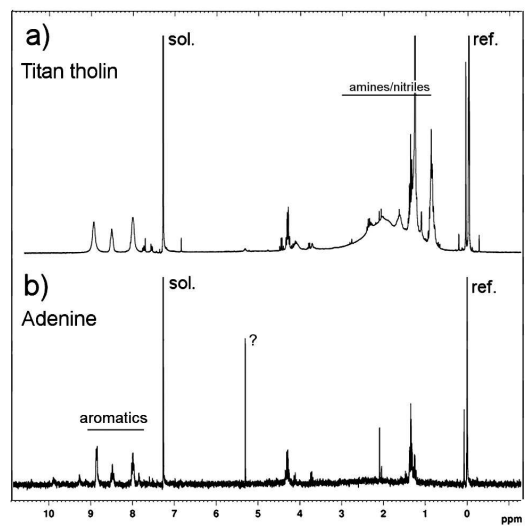}
\caption{$^1$H NMR spectra obtained after the derivatization process of the
Titan tholin produced by soft X-rays (a) and adenine standard (b).} \label{fig-NMR}
\end{figure}

A second aliquot of the derivatized Titan tholin at room temperature was analyzed by $^1$H NMR. The spectra obtained for both the tholin and the adenine standard, results in the same values of chemical shifts for the aromatic hydrogens (7 - 9 ppm), evidencing the formation of adenine in the experiments using synchrotron irradiation and confirming the previous analysis by gas-chromatography. The $^1$H NMR spectra obtained after the derivatization process of the Titan tholin produced by soft X-ray and adenine standard is shown in Fig.~\ref{fig-NMR}a and \ref{fig-NMR}b, respectively.

Despite of the infrared spectra of organic residue have presented small features closer to the adenine infrared bands (see Fig.~\ref{fig-FTIR}b), adenine was only effectively detected after chromatographic and NMR analysis of the organic residue, done at room temperature. To verify if adenine it self, and not its precursor species, was indeed directly produced by soft X-ray photolysis, more experiments are needed. This issue will be the subject of future investigation with the employing of a high resolution time of flight spectrometer coupled to the vacuum chamber which allows accurate \emph{in-situ} analysis of surface chemistry.

\section{Discussion}

As discussed by Basile and collegues\cite{basile84}, nitrogen-ring compounds (e.g. purines and pyrimidines) have been observed from the processing (mainly by thermal heating and electric discharges) of non-biological matter (e.g. primitive Earth atmosphere analogs) along the last 40 years. The first evidence for the possibility of prebiotic synthesis of adenine was presented by Oró\cite{Oro60,Oro61} and by Oró and Kimball\cite{OroKimbal61} from thermal heating of ammonium cyanide solution and HCN pentamerizaton. A detailed mechanism of adenine synthesis from HCN-pentamers was given elsewhere\cite{Glaser07}. Lowe and coworkes\cite{Lowe63} have shown that the heating of a solution containing hydrocyanic acid with aqueous ammonia for a long time (90$^{\circ}$C for 18 h) also produce adenine and other nitrogen-ring compounds. Adenine was also observed after the heating of solution of hydrogen cyanide in liquid ammonia for an extended period of time at elevated temperatures\cite{wakamatsu66, Yamada68}.

One of the first experiments employing electrons, in which adenine was observed, was performed by Ponnamperuma and collegues\cite{Ponnamperuma63}. In this experiment a gaseous and liquid mixture of methane and ammonium hydroxide, NH$_4$OH, was bombarded by 4.5 MeV electrons. Since the ammonium hydroxide was obtained from the mixture between ammonia and water, this investigation have clearly established adenine as product of the irradiation of methane, ammonia and water. Khare et al.\cite{Khare84} have simulated a Titanian-like atmosphere by irradiation with high energy electrons in a plasma discharge. In their work,  among the hundred compounds identified by the chromatographic analysis, including several amino acids, aliphatic and aromatic nitriles and nitrogenous rings, there were more than thirty nitrogenous rings including adenine. Other synthesis of purine and pyrimidine bases, and related compounds by the action of electric discharges on a primitive atmosphere analog containing CH$_4$, NH$_3$ (or N$_2$), and H$_2$O have been reported elsewhere\cite{Yuasa84, Kobayashi86a, Kobayashi86b}.

Chromatographic analysis performed by Pietrogrande and coworkers\cite{Pietrogrande01} on the organic residue produced by corona discharges into Titan atmosphere analog have indicated the presence of several cyclic and aromatic compounds containing nitrogen like pyrimidine (C$_4$H$_4$N$_2$), pyridine (C$_5$H$_5$N), 1H-pyrrole (C$_4$H$_5$N) and benzonitrile (C$_7$H$_5$N) but no adenine was observed. Recently, McGuigan et al.\cite{McGuigan06} have performed an analysis of Titan tholin pyrolysis products by comprehensive two-dimensional gas chromatography-time-of-flight mass spectrometry. Despite the observation of several nitrogen compounds including benzonitrile (C$_7$H$_5$N), methyl-benzonitrile, pyrrole (C$_4$H$_5$N), alkyl subistitute pyrroles, alkyl-substituted benzenes and polycyclic aromatic hydrocarbons (PAHs) compounds, no adenine or other nucleobases were observed.

Many nitriles and nitrogen-containing heterocyclic compounds have also been observed from experiments employing ion bombardment on simple gas mixtures simulating planetary atmospheres. Kobayashi and Tsuji\cite{Kobayashi97} have showed that uracil, one of the four RNA bases, can be formed from simulated primitive atmosphere composed of CO, N$_2$ and H$_2$O, by proton irradiation. Yamanasi and collegues\cite{Yamanashi01} have identified cytosine in the hydrolysate product obtained after the proton irradiation of a mixture of CO, NH$_3$ and H$_2$O. Thymine have been identified among the proton irradiation products of a mixture of CH$_4$, CO and NH$_3$ by Kobayashi and coworkers\cite{Kobayashi04}. However in these experiments adenine was not observed.

To our knowledge the present work is the first experiment employing photons in which adenine was synthesized in an primitive atmosphere simulation. The interaction between soft X-rays and matter produce energetic secondary electrons which could be essential for  adenine synthesis, since adenine have been observed in previous experiments involving discharges and electron bombardment as discussed before\cite{Ponnamperuma63, Khare84, Kobayashi86a}.

In this work no amino acids or other nucleobases (guanine, cytosine, uracil or thymine) were observed in the residues from chromatographic or NMR analysis. This could be attributed by the small radiation resistance of theses species to soft X-rays\cite{Pilling2008}. Recently, Pilling and collegues\cite{Pilling2009} have observed from experiments involving the photodegradation solid phase and gas-phase biomolecules by soft X-rays, that adenine is at least 10 times less radiation sensitive than uracil and thousand times more resistent than amino acids. In addition, following the molecular orbital calculations performed by Pullman \& Pullman\cite{Pulman62,Pulman60}, of all the biologically important nitrogen-ring compounds like purines and pyrimidines, adenine has the greatest resonance energy. This makes its not only more likely to be synthesized but also confer radiation stability upon its. From the previous statements we suggest that during the continuous exposure to soft X-rays the possible formed amino acids or its precursors are fully dissociated (processed) by ionizing radiation. A similar result was observed by Ponnamperuma et al.\cite{Ponnamperuma63} in experiments involving the irradiation of gaseous and liquid CH$_4$ and ammonium hydroxide NH$_4$OH mixture by 4.5 MeV electrons.

As pointed out by Ponnamperuma et al.\cite{Ponnamperuma63}, the apparent preference for adenine synthesis may be also related to adenine´s multiple roles in biological system. Adenine, besides to be a constituent of both DNA and RNA, is also a unit of many important cell cofactors (e.g. ATP, ADP, DPN, TPN, FAD and coenzime A).

Recently measurements done by Cassini spacecraft have revealed that Titan is not in synchronous rotation (same face to the planet) with Saturn indicating a possible internal ocean of liquid water\cite{Lorenz08}. If Titan have experienced a warm period in the past, promoted by external (e.g. comets impacts, Saturn magnetic field and tide effects) or internal (e.g. vulcanism, intense radioactive decay) forces to make liquid its water-ammonia ices some prebiotic molecules precursors could have been hydrolyzed and a primitive life could have had a chance to flourish there. In the other hand, in the very far future, when the sun becomes a magnificent red-giant and fill the solar system up to Earth orbit, Titan land surfaces may changes to liquid-land surfaces allowing these prebiotic compounds, produced and processed by radiation and energetic particles over a billions years, to react. When this time comes, life will have another chance to arise like happened in the primitive Earth.

\section{Conclusion}

In this work we present the chemical alteration produced by the interaction of soft X-rays (and secondary electrons) on Titan aerosol analogs. The experiments simulates roughly $7\times 10^6$ years of solar soft X-ray exposure on Titan atmosphere. The presence of small amount of water and CO in the sample simulates periods of heavy cometary bombardment on this Saturn´s moon. \emph{In-situ} infrared analysis have shown several organic molecules created and trapped in the ice at $\sim$ 15 K including the reactive cyanate ion ONC$^-$, nitriles and possibly amides and esters. Thermal heating of frozen tholin changes drastically its chemistry resulting in a organic residue rich with C-C and C-N aromatic structures. In Titan, the processed aerosols will be deposited along the time at the surface and/or at the bottom of lakes/rivers, leaving with them newly formed organic species.

Gas chromatography and H$^1$ NMR analysis of the organic residue at room temperature have shown that among several nitrogen-compounds, adenine, one of the DNA-nucleobases, is one of the most abundant species produced due to irradiation by soft X-rays. This confirm previous studies suggesting that the organic chemistry onto Titan atmosphere and surface should be complex, being rich in pre-biotic molecules like adenine and amino acids (or/and its precursors species). Molecules like these in the early Earth have found a place that allows life (as we know) to flourish, a place with liquid water.
\\

\textbf{The final version of this manuscript can be obtained at \texttt{http://pubs.acs.org/doi /abs/10.1021/jp902824v}}\\
\vspace{0.5cm}

\footnotesize \textbf{Acknowledgement}
The authors would like to thank the staff
of the Brazilian Synchrotron Facility (LNLS) for their valuable
help during the experiments. We are particularly grateful to Dr. A. M. Slovic, Dr. F. C. Vicentin, Dr. P. T. Fonseca, Dr. G. Kellerman, Dr. L. Ducatti and MSc. F. R. Francisco for technical support. The authors also thanks Dr. C. M. da Conceição for the critical reading of the manuscript. This work was supported by LNLS, CNPq and FAPESP. \normalsize


\newpage

\end{document}